\documentclass[journal]{IEEEtran}
\usepackage[hyphens]{url}
\usepackage{graphicx}    
\usepackage{subcaption}  
\usepackage{float}  
\usepackage{ifthen}
\usepackage{tcolorbox} 
\usepackage{framed}    
\usepackage{mdframed} 
 \usepackage{booktabs}
 \usepackage{amsmath} 
 \usepackage{diagbox}
 \usepackage[hidelinks]{hyperref}

\begin{document}

\title{Beyond the Voice: Inertial Sensing of Mouth Motion for High-Security Speech Verification}

\author{Ynes~Ineza, Muhammad~A.~Ullah, Abdul~Serwadda, and Aurore~Munyaneza%
\thanks{The authors are with the Department of Computer Science, Texas Tech University, Lubbock, TX 79409 USA (e-mail: yineza@ttu.edu, mazizull@ttu.edu, amunyane@ttu.edu, abdul.serwadda@ttu.edu).}
}

\maketitle

\begin{abstract}

Voice interfaces are increasingly used in high-stakes domains such as mobile banking, smart-home security, and hands-free healthcare. Meanwhile, modern generative models have made high-quality voice forgeries inexpensive and easy to create, eroding confidence in voice authentication alone. To strengthen protection against such attacks, we present a second authentication factor that combines acoustic evidence with the unique motion patterns of a speaker’s lower face. By placing lightweight inertial sensors around the mouth to capture mouth opening and evolving lower-facial geometry, our system records a distinct motion signature with strong discriminative power across individuals. 

We built a prototype and recruited 43 participants to evaluate the system under four conditions: seated, walking on level ground, walking on stairs, and speaking with different language backgrounds (native vs. non-native English). Across all scenarios, our approach consistently achieved a median equal-error rate (EER) of 0.01 or lower, indicating that mouth-movement data remain robust under variations in gait, posture, and spoken language. We discuss specific use cases where this second line of defense could provide tangible security benefits to voice authentication systems.
\end{abstract}

\section{Introduction}

Voice-based deepfakes have recently emerged as potent weapons in social engineering attacks, corporate fraud, and impersonation schemes. From fraudulent phone calls to fake voicemail messages, synthetic voices are increasingly integral to the cybercriminal toolkit \cite{bunn2023artificial,voice-cloning-2025}. A key factor driving this rise of voice deepfakes is the ready availability of voice data for training generative models. In today’s digital world, many individuals have hours of their speech accessible online via podcasts, interviews, conference talks, social media videos, and even corporate earnings calls. These plentiful audio samples provide the raw material for Artificial Intelligence (AI)-driven voice cloning. 

Indeed, with only a short snippet of someone’s voice, scammers can train a deepfake model to generate convincing utterances in that person’s exact vocal timbre. For example, the Eleven labs platform requires a minimum of just 30 minutes of audio to generate a high quality voice clone \cite{eleven-labs}. This ease of data collection and model training has caused voice-based impersonation attacks to accelerate at a worrying pace.

A variety of countermeasures have been proposed to combat these threats. One approach is to watermark AI outputs so that synthetic content can be detected by spotting the watermark. However, this method fails when considering the many open-source models that do not enforce watermark standards. Another suggestion is to build AI detectors that exploit subtle ``fingerprints'' unique to synthetic audio. These tools, however, risk obsolescence whenever more advanced forgery techniques appear. A third approach involves digitally signing legitimate recordings to distinguish them from fake ones, but this requires extensive infrastructure and certificate management, which is not feasible at scale.

Since none of these ideas alone fully address the underlying problem, this paper introduces a complementary strategy: adding a second modality to basic voice authentication. The underlying logic is straightforward: if an adversary must forge two data streams (e.g., audio + physical signals) instead of just one, the barrier to impersonation rises considerably. Crucially, this secondary modality should be something not typically published in the public; unlike voices, which are widely exposed and can be scraped by attackers.  

In this paper, we propose that, for certain high-security use-cases, inertial measurements of the mouth during speech offer such a modality. Our core hypothesis is that humans exhibit distinct jaw lengths, muscle structures, and mouth movements when articulating words, leading to unique sensor signatures that augment standard voice data. 
By capturing fine-grained mouth-motion dynamics in tandem with the spoken audio, a system could more confidently verify that the speaker is legitimate rather than a deepfake simulation.
 
We prototype this concept and conduct rigorous experiments to illustrate its feasibility. Below, we summarize our principal contributions:

{\bf{(1) Mouth Kinematics as a Biometric:}}
We propose a novel biometric authentication scheme that captures individualized mouth movement patterns using multiple inertial sensors strategically positioned around the mouth region. We design and implement a functional prototype and evaluate it through experiments involving participants speaking freely while wearing the device. We also identify potential applications, some of which having similar headgear already in use, enabling the seamless integration of our system.

{\bf{(2) Comprehensive Evaluation Across Realistic Conditions and Speaker Populations:}}
We rigorously evaluate our authentication system using data from 43 participants across a range of conditions, including sitting, walking on flat ground, and walking on stairs. From the collected signals, we extract a broad set of time-domain and frequency-domain features, assess their discriminative power, and select a compact feature set for classification. We compare traditional machine learning classifiers such as SVM with neural network-based approaches like Long Short-Term Memory (LSTM). The LSTM consistently performs best, achieving median equal error rates (EERs) below 0.01. 

Our evaluation includes both native and non-native English speakers, allowing us to assess the system’s generalizability across diverse language backgrounds. We also analyze the distribution of EERs across individuals and examine the impact of sensor placement on biometric distinctiveness. These insights are critical for informing future designs and optimizing sensor configurations.

{\bf(3) Video-Driven Attack Evaluation:}
We develop and rigorously test a {\it video-based} impersonation attack, the most realistic threat an adversary faces when our inertial data are not publicly available.  
An attacker crawls readily available footage of the target (e.g., interviews, vlogs, conference talks) and applies state-of-the-art face-tracking to recover 3-D trajectories of landmarks that coincide with our sensor locations.  
By taking second-order temporal derivatives of these trajectories, we synthesize per-frame accelerations in the \(X\), \(Y\), and \(Z\) directions to mimic the inertial signals captured by our prototype.  

Because Internet videos appear at widely varying resolutions and frame rates, we repeat the extraction at multiple frame rate and resolution settings to emulate realistic quality constraints before submitting the resulting traces to our trained classifier.  
This evaluation quantifies how well inertial mouth-movement biometrics withstand a resourceful adversary armed only with public video, providing a rigorous end-to-end measurement of this threat.

\begin{figure*}
    \centering
    \includegraphics[width=.80\linewidth]{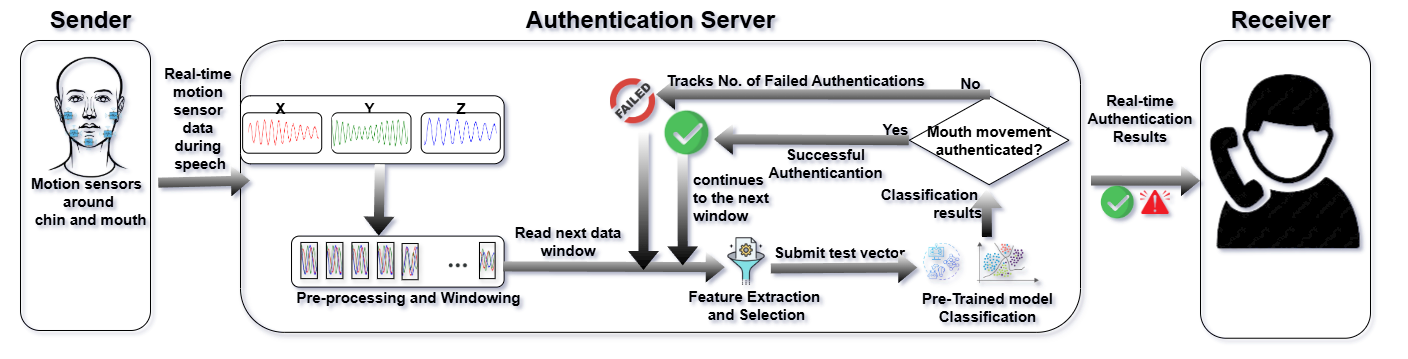}
    \caption{Operational flow of continuous authentication using mouth-motion signals}
    \label{ccs_flowdiagram}
\end{figure*}



\section{System and Threat Model}
\label{threat1}

\subsection{System Context}

We consider a scenario in which voice authentication, whether by human listeners (for example, during a phone call) or by automated verification tools, happens at the same time as real-time mouth movement sensing. The user wears a set of inertial sensors around the mouth that capture physical movement data while they speak. This inertial sensor data is then evaluated alongside the audio.

Our experiments assume that attackers already possess near-perfect ability to forge the user’s voice, given the abundance of publicly available speech recordings and modern deepfake methods. We therefore treat voice forgery as a given and do not implement it in our work since prior research has extensively explored that topic \cite{pianese2024training, amezaga2022availability, gonzalez2023enhancing}. Instead, we focus on the edge case, where mouth motion data serves as the last line of defense, under the rationale that an attacker who has successfully spoofed the voice would also need to replicate the user’s personalized mouth movements to succeed. By verifying these real-time mouth dynamics, the system significantly raises the barrier for impersonation attempts, because mouth kinematics as measured by inertial sensors, would typically not be exposed in the public domain like voices are. 

Figure~\ref{ccs_flowdiagram} illustrates how the mechanism proposed in this paper could be deployed in practice. In this design, inertial mouth-motion data is continuously evaluated, possibly by a trusted server, as the user speaks. Authentication failures generate warning triggers but do not immediately terminate the communication session. This approach reduces the risk of false negatives, which are common in behavioral biometric systems, from prematurely disrupting legitimate conversations. At the same time, it allows the receiver to monitor the frequency and severity of authentication anomalies and make informed decisions. If the pattern of failures suggests a voice forgery, the receiver may choose to terminate the conversation, delay any actions prompted by the speaker, or initiate additional verification steps before proceeding.

\subsection{Type of Speech Under Consideration}
We focus on a ``free speech'' scenario, which means the user’s mouth movement signals are captured while they speak any arbitrary content, rather than reading from a fixed script or repeating pre-determined phrases. The assumption is that each individual’s physical jaw structure, muscle tension, and habitual speaking mannerisms produce distinctive patterns, such as mouth opening extent, jaw speed, head tilt, and characteristic pauses between words. Our hypothesis is that these personalized motion features remain relatively stable across different spoken content.

This focus on free speech has practical implications for real-world use, because it allows our biometric to operate on any speech the user produces, without requiring them to say specific words during enrollment or authentication. In other words, it is not necessary to map every possible spoken utterance to known jaw movements in the training dataset. Instead, we collect representative training data from the user as they speak naturally. Later, at authentication time, the system can track new speech from the same user and compare the resulting motion signals to the user’s baseline profile.

\subsection{Use-Cases and Example Applications}

Our approach assumes high-security voice authentication scenarios in which users are able and willing to wear a head-mounted sensor system during critical communications. In many professional settings, some form of headgear is already a routine part of operations. For example, military personnel often wear helmets with integrated radios, so adding jaw sensors would impose minimal overhead. A forged voice in a battlefield context could lead to false orders that jeopardize missions or put troops at risk. Similarly, air traffic controllers rely on specialized headsets to coordinate flights in real time, and a convincing impersonation of a controller’s voice could result in disastrous flight instructions. Equipping these headsets with jaw-motion sensors adds an extra layer of security, making it much harder for an attacker to fool the system with synthetic speech alone.

High-stakes financial calls provide another environment where preventing unauthorized impersonation is paramount. Executives or bankers discussing large transactions can wear sensor-embedded caps to help ensure that only genuine participants are recognized. The same logic applies to other critical communications, such as official government lines or emergency response centers, where accurately confirming a speaker’s identity is vital for safety and effectiveness. In these contexts, wearing a sensor-based device is not out of place, or at least should be considered acceptable, given the risks of voice impersonation and the importance of reliable verification.

\subsection{Attacker's Capabilities}

As previously described, we assume a scenario where the attacker has successfully cloned the user's voice and is thus only thwarted by our jaw-movement biometric layer. Crucially, the attacker does not have access to the victim's sensor data, meaning they cannot directly craft a precise forgery of the jaw movement pattern. Within these constraints, we define two realistic adversarial strategies.

In the first strategy, termed the \textbf{zero-effort attack}, the attacker attempts to bypass our defense by submitting jaw movement sensor data from other individuals, rather than from the intended victim. In practice, the adversary would physically wear similar sensors on their own jaw while speaking, and submit these real-time captured patterns as if they belonged to the targeted user, in hopes of fooling the authentication system.

The second strategy is a \textbf{video-based attack}, where the attacker identifies publicly available footage of the victim speaking such as interviews, presentations, or social media videos. The attacker then uses video tracking techniques to estimate the motion at the locations corresponding to our sensor placement. From these visual cues, the attacker reconstructs synthetic inertial sensor data that approximates the victim’s jaw dynamics and attempts to inject this into the authentication system.

\begin{figure*}
    \centering
    \begin{subfigure}[b]{0.18\textwidth} 
        \centering
        \includegraphics[width=\linewidth]{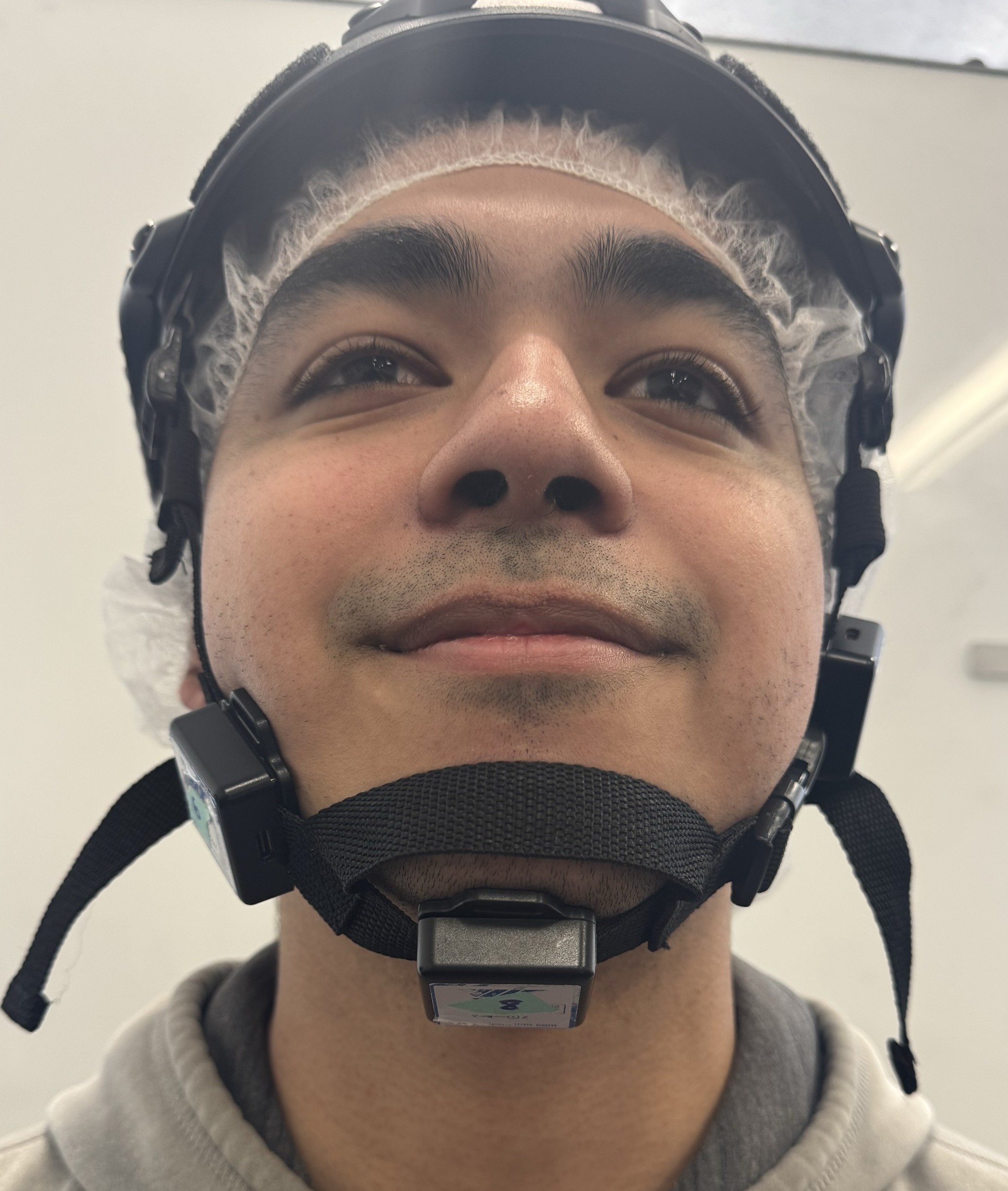}
        \caption{Sensor below the chin \\}
        \label{fig:sub1}
    \end{subfigure}
    \hfill
    \begin{subfigure}[b]{0.18\textwidth}
        \centering
        \includegraphics[width=\linewidth]{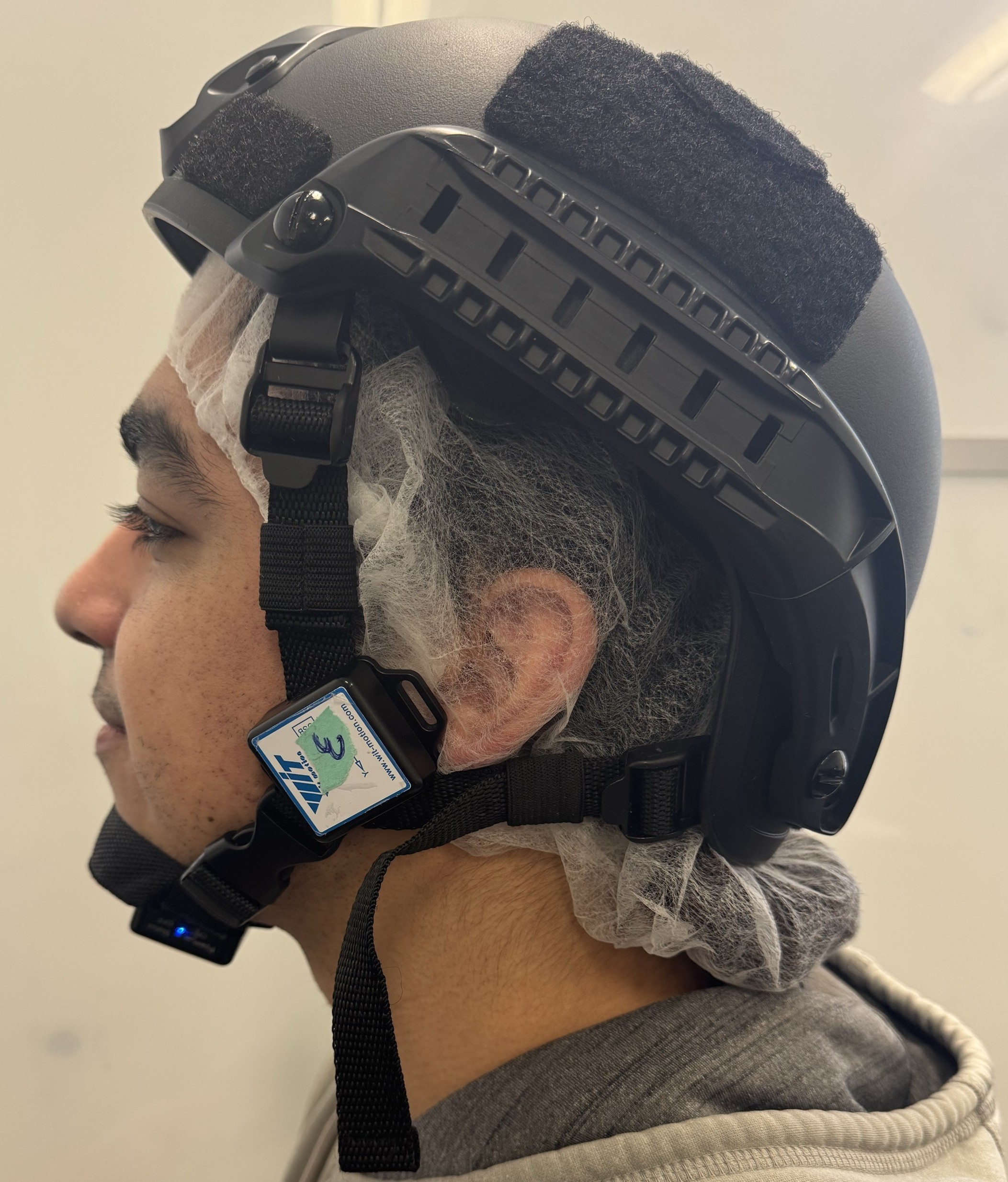}
        \caption{Sensor on left upper jaw}
        \label{fig:sub2}
    \end{subfigure}
    \hfill
    \begin{subfigure}[b]{0.18\textwidth}
        \centering
        \includegraphics[width=\linewidth]{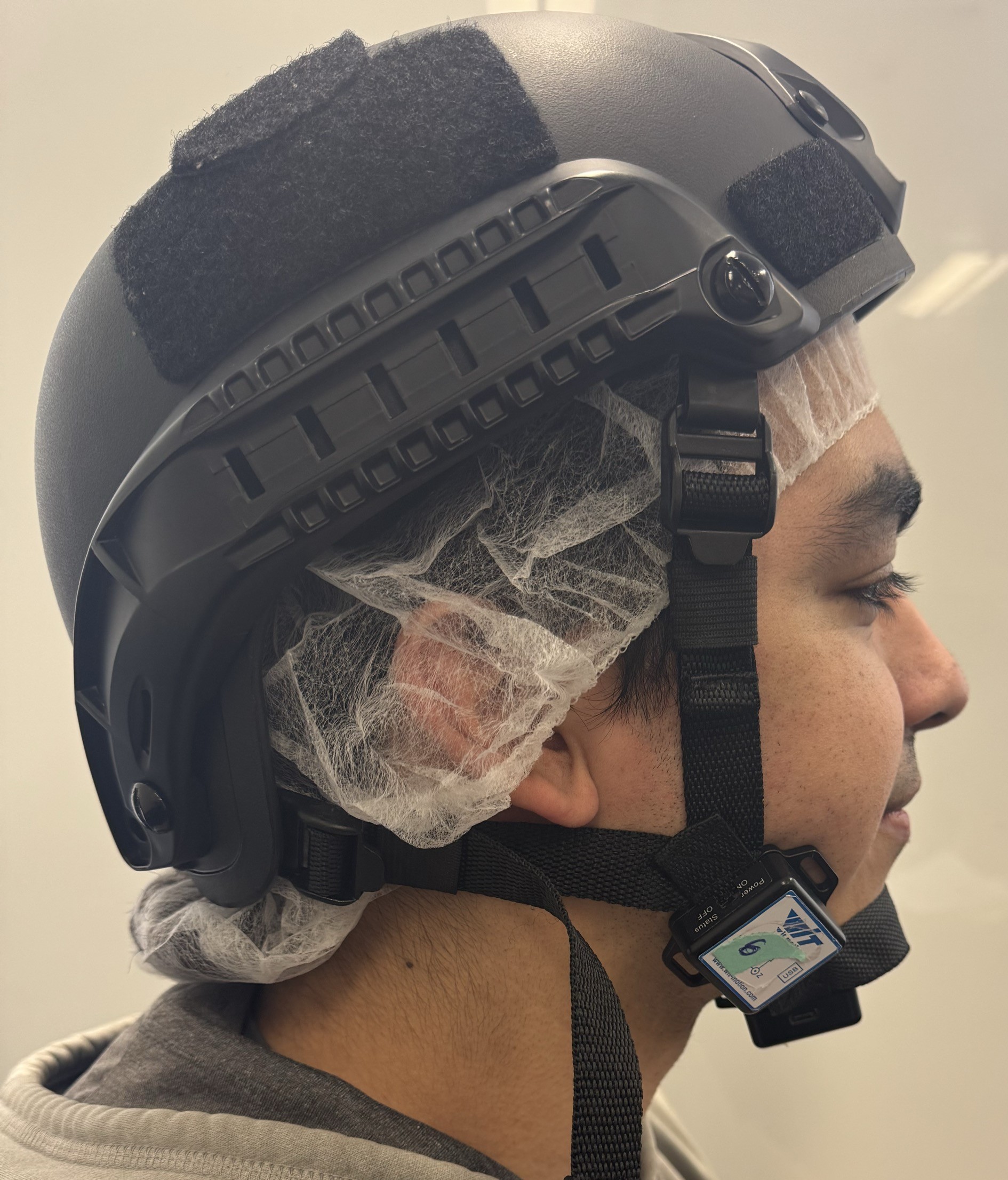}
        \caption{Sensor on right lower jaw}
        \label{fig:sub3}
    \end{subfigure}
    \caption{User wearing our authentication prototype}
\label{prototype}
\end{figure*}
\section{Mouth Movement Authentication: Design Rationale and Preliminary Experiments}
\subsection{Overview}
Although the full production of speech involves many internal mechanisms such as vocal cords and tongue movement, external mouth movement (specifically around the lower jaw, cheeks, and lips), also plays a significant role in how people articulate words. When a person speaks, the lower jaw frequently moves up and down, sometimes shifting forward or laterally for certain phonemes, while the upper jaw remains relatively fixed. Meanwhile, facial skin near the cheeks and lips flexes inward or outward as the mouth opens and closes. Our core hypothesis is that these external motion patterns vary from individual to individual due to differences in skeletal geometry, muscle configuration, and habitual speaking style.

The fundamental goal of our design is to capture these personalized kinematics using strategically placed inertial sensors around the mouth region. In this section we provide the rationale behind our design, and present results from preliminary experiments that we undertook to gain some high-level insights into the potential of our hypothesis before we conducted the fully-fledged study.

\subsection{Sensor Placement Rationale in Prototype}
To explore the discriminatory motion signatures generated during natural speech, we incorporate three distinct sensor placement configurations in our experimental design. These locations—distributed across the lower face—capture vertical displacement, lateral skin stretch, and geometric variation associated with jaw articulation. Section \ref{implementation} provides additional detail on the sensing hardware, strap interface, and implementation used to operationalize these configurations.
\subsubsection{Chin Sensor (Capturing Vertical Mouth Displacement)}
Positioning a sensor directly beneath the chin allows us to measure the vertical travel of the mouth as the speaker forms vowels, consonants, and transitions between them. The chin’s up-and-down motion is the most direct indicator of mouth opening magnitude. Because each individual’s lower jaw size, muscle tone, and bone structure are different, we hypothesize that the speed, depth, and angle of this motion are also unique. By recording these signals in real time, we obtain a core component of the user’s mouth-opening signature.

\subsubsection{Side Sensors (Tracking Skin Stretch and Lateral Motion)}
The skin along the left and right sides of the jaw often stretches and contracts when a speaker forms different phonemes, particularly those requiring wide mouth openings or specific lip movements. To capture these lateral expansions and subtle vibrations, we attach two sensors on each side of the jaw area. One sensor is placed slightly higher (near the upper jaw region), while the other is placed lower (closer to the jawline itself). This vertical offset ensures that each sensor picks up a somewhat different pattern of skin tension, strap tension, and rotational movement. In effect, we measure not only how far the mouth opens but also how the cheek and jawline shift across multiple degrees of freedom.

\subsubsection{Lower-Face Geometry Through Strap Coupling}
Beyond capturing mouth movement itself, each sensor also helps trace the lower face region’s distinct geometry. The accelerations and angles measured on the left and right sides depend on how broad or narrow the cheeks are, how prominent the jawbone is, and other anatomical factors such as muscle tone or overall facial structure around the mouth. Individuals can differ significantly in these dimensions. Some have a narrower lower face, while others are more rounded, leading to unique inertial signals for the same mouth movement. In order to maintain a consistent mapping between sensor orientation and the user’s lower-facial contours, we rely on a flexible strap system that is configurable to keep each sensor firmly against the skin. This tight coupling ensures that the sensors move in unison with the user’s lower face, rather than slipping or shifting in ways that might distort the recorded motion patterns.

\begin{figure}[b]
    \centering
    \includegraphics[width=0.3\linewidth]{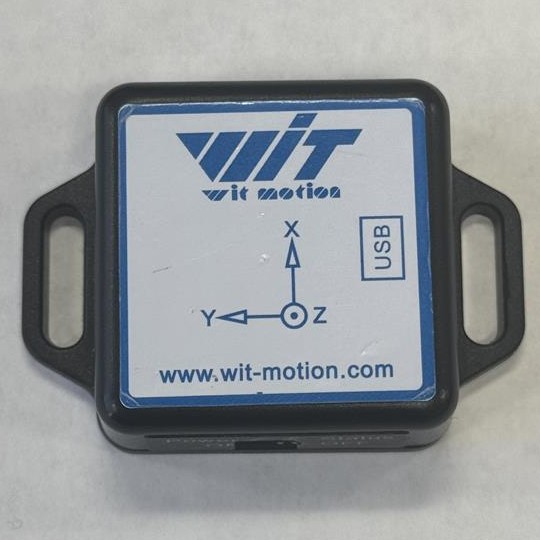}
    \caption{Motion sensor used in our experiment}
    \label{sensor_image}
\end{figure}

\subsection{Prototype Implementation}
\label{implementation}
We used an Airsoft Tactical Helmet \cite{helmet,novritsch_helmet}, which features a robust yet flexible strap system that can be easily tightened or loosened to fit different head sizes. Such a configuration is frequently seen in military and tactical helmets, sports safety gear, and certain industrial headgear.

We attached Witmotion WT901BLECL sensors \cite{sensors} at three positions along these straps near the jaw. Figure \ref{sensor_image} shows one of these sensors. Each unit includes a built-in three-axis accelerometer, gyroscope, magnetometer, and angle measurement capabilities, and transmits real-time motion data via Bluetooth to a Windows 10 PC for data collection and visualization. The sensors operate at the rate of 100~Hz on average.

Figure \ref{prototype} shows a user wearing our prototype. The polypropylene layer shown between the cap and the user’s head is not part of the prototype itself; we used it for hygienic reasons, since multiple participants wore the same helmet during our experiments. In a practical deployment, these sensors could be embedded directly into a helmet or strap lining, making them far less obtrusive. Smaller, more streamlined sensor modules would allow additional placement options, potentially providing finer-grained coverage of jaw movement across more regions of the mouth. 

\begin{figure*}[ht]
    \centering
    \begin{subfigure}[b]{0.25\textwidth} 
        \centering
        \includegraphics[width=\linewidth]{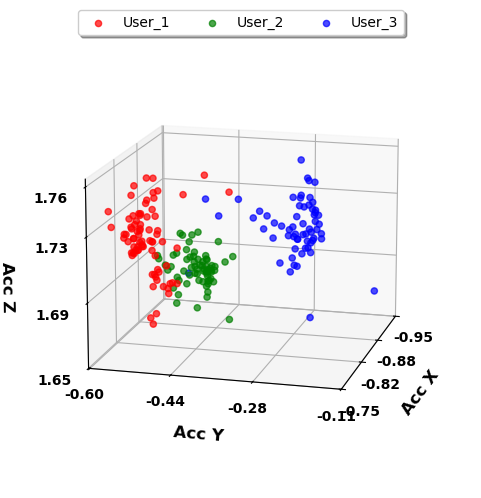}
        \caption{Upper Left Jaw}
        \label{fig:sub1}
    \end{subfigure}
    \hfill
    \begin{subfigure}[b]{0.25\textwidth}
        \centering
        \includegraphics[width=\linewidth]{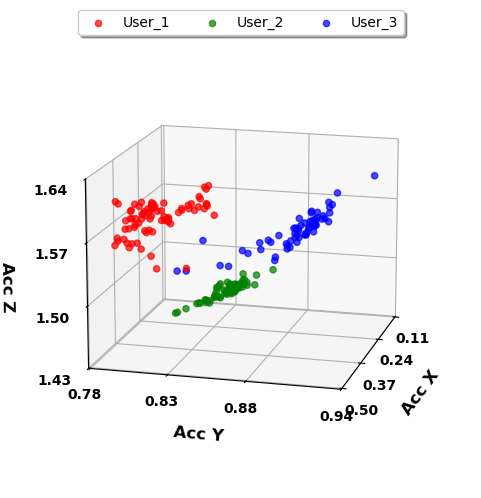}
        \caption{Lower Right Jaw}
        \label{fig:sub2}
    \end{subfigure}
    \hfill
    \begin{subfigure}[b]{0.25\textwidth}
        \centering
        \includegraphics[width=\linewidth]{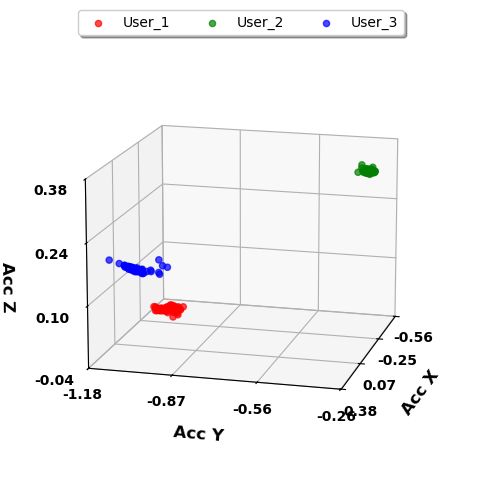}
        \caption{Below the Chin}
        \label{fig:sub3}
    \end{subfigure}
    \caption{Mean acceleration in the X, Y and Z directions for 3 users speaking continuously for 30 seconds. Each of the users spoke different content from the other users. }
    
    \label{preliminary-results-different}
\end{figure*}

\begin{figure*}[ht]
    \centering
    \begin{subfigure}[b]{0.25\textwidth} 
        \centering
        \includegraphics[width=\linewidth]{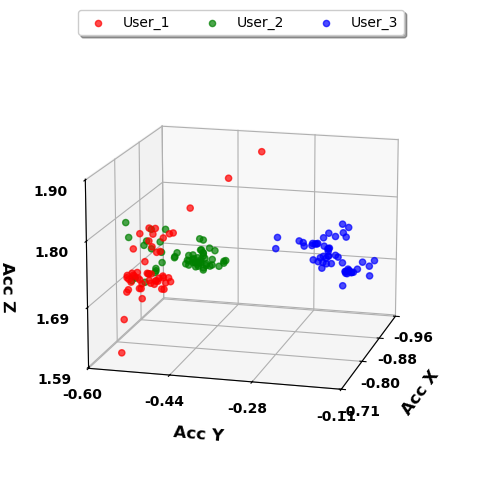}
        \caption{Upper Left Jaw}
        \label{fig:sub1}
    \end{subfigure}
    \hfill
    \begin{subfigure}[b]{0.25\textwidth}
        \centering
        \includegraphics[width=\linewidth]{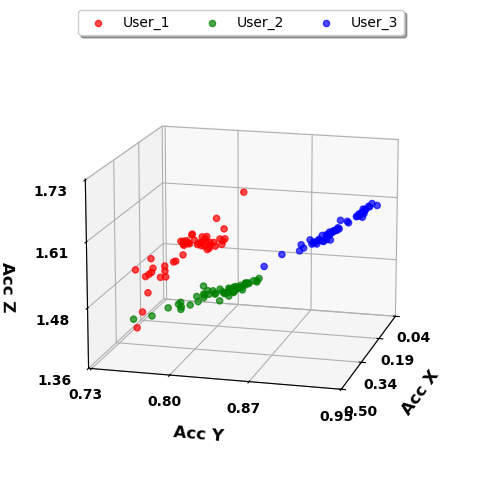}
        \caption{Lower Right Jaw}
        \label{fig:sub2}
    \end{subfigure}
    \hfill
    \begin{subfigure}[b]{0.25\textwidth}
        \centering
        \includegraphics[width=\linewidth]{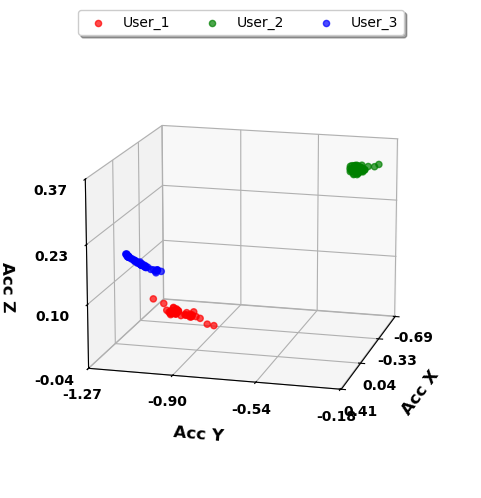}
        \caption{Below the Chin}
        \label{fig:sub3}
    \end{subfigure}
    \caption{Mean acceleration in the X, Y and Z directions for 3 users who read from a common script. All users spoke exactly the same content.}
    
    \label{preliminary-results-same}
\end{figure*}

\subsection{Preliminary Experiments}
We conducted a series of small-scale experiments to see whether our mouth-movement sensor configuration could indeed produce user-specific patterns. These preliminary tests offered a high-level assessment of the system’s potential, helping us decide how to proceed toward a more comprehensive design and in-depth evaluation. Below, we detail these experiments and their outcomes. 

In the first exploratory experiment, we recruited a small group of volunteers who wore our prototype sensor device while speaking freely for approximately 30 seconds. During this period, we recorded sensor data capturing jaw motion, head movement, and mouth-opening dynamics. From these raw signals, we extracted several basic features and visualized their combinations in three-dimensional feature spaces.

Figure \ref{preliminary-results-different} presents the results of this experiment involving three users. Accelerometer data from three distinct sensor locations around the jaw were segmented into 0.5-second windows, and the mean values of each window were computed separately along the X, Y, and Z axes. The resulting plots illustrate distinct clusters for each user at all three sensor locations, indicating clear visual separability. This initial evidence suggested that a more comprehensive and rigorously selected set of features, combined with advanced classification algorithms, might indeed support a robust biometric authentication modality.

One potential limitation of the free-speech scenario depicted in Figure \ref{preliminary-results-different} is that observed differences might partially result from variability in speech content rather than purely biometric traits. To address this, we conducted a complementary experiment in which participants read the same scripted text, standardizing the speech content across users. By controlling for linguistic variability, we aimed to attribute any observed differences more confidently to inherent anatomical and kinematic traits of each user’s jaw movements. Using similar data processing steps and visualization methods, Figure \ref{preliminary-results-same} shows the results from this second experiment. Notably, even with identical spoken content, the visualizations again revealed distinct clustering patterns unique to each user.

Collectively, these preliminary experiments, despite their limited scale, provided initial but encouraging evidence that jaw motion data capture inherent biometric traits robust enough to persist across varied speaking conditions. The free-speech experiment closely reflects real-world application scenarios, while the scripted-speech trial helps isolate anatomical differences from variation in language. Taken together, these findings encouraged us to proceed with a more extensive collection effort and a systematic evaluation framework, which we describe in the following sections.
 
\section{RELATED WORK}

Our work relates to two primary research directions. The first involves methods that use head-worn or ear-worn devices to detect subtle vibrations that are then used to augment speech authentication. The second direction comprises of the broader family of techniques proposed to combat fake audio.  

\subsection{Approaches Based on Ear-Worn and Head-Worn Devices}
\label{head-worn}
The Face-Mic system, proposed by Shi et al.~\cite{shi2021facemic}, demonstrates that the motion sensors embedded in AR/VR headsets can capture speech-related facial dynamics and head movements, even in the absence of a microphone. These headsets, which rest above the nose, are able to record subtle skin and bone vibrations during speech. In experiments involving 45 participants using four commercial VR headsets, Face-Mic achieved high accuracy in speaker and gender identification and showed promising results in reconstructing speech content. Although introduced as a privacy-focused attack, the system remains relevant to our work by virtue of using speech-induced vibrations for identity inference.

Several other works have used inertial sensors embedded in earphones (i.e., “earables”) for user authentication. The MandiPass system, first introduced in a conference paper~\cite{liu2021mandipass} and later extended in a journal version~\cite{liu2022secure}, uses an inertial measurement unit (IMU) integrated into a standard earphone to capture jawbone vibrations during speech. In evaluations involving 34 users speaking digits 0 through 9, MandiPass achieved equal error rates (EERs) below 3\% using a single IMU positioned on the right ear.

Jawthenticate~\cite{jawthenticate} also falls into this category of ear-based sensing but adopts a dual-sensor approach. It places one sensor behind the ear to capture head motion and another several centimeters forward, near the jaw, to lightly touch the face and measure speech-related facial vibrations. In experiments with 41 participants, the system achieved a balanced accuracy (BAC) of 97.07\% using only 3 seconds of speech. It was shown to be robust across multiple speech contents and even worked under conditions of inaudible speech articulation.

VAuth~\cite{feng2017vauth} takes a different form, offering continuous voice-user authentication using body-surface vibrations captured by accelerometers embedded in everyday wearable objects such as earbuds, eyeglasses, or necklaces. VAuth compares these vibrations against the spoken signal captured by the microphone and only executes commands when the signals align. In a study involving 18 participants issuing 30 different commands, VAuth achieved near-perfect accuracy across the three wearable form factors.

\vspace{1ex}
We highlight three key distinctions between our work and the prior literature:

\paragraph{Biometric Trait Being Measured.}  
To our knowledge, our system is the first to directly and comprehensively measure the spatial and temporal dynamics of the mouth and lower face during speech. A centrally placed sensor beneath the chin captures vertical mouth movement, including the degree of opening and closing. Two side sensors, placed on the left and right sides of the mouth, record lateral skin and muscle motion. These signals come from the core articulators used in speech and reflect the physical details of how a person talks.

In contrast, previous systems rely on indirect signals collected from other parts of the face or head. Face-Mic~\cite{shi2021facemic} uses motion data from a VR headset positioned above the nose, an area not involved in articulation. MandiPass~\cite{liu2021mandipass, liu2022secure} and Jawthenticate~\cite{jawthenticate} collect signals near the right ear, which only pick up broad jaw movement. VAuth~\cite{feng2017vauth} uses sensors on accessories like earbuds, glasses, or necklaces, and does not measure any part of the mouth or lower face.

The lower face, especially the mouth, plays a central role in speech production. Movements in this region are shaped by each individual’s muscle coordination and speaking habits, making them valuable for distinguishing between users. By capturing these movements directly, our system provides the first clear evidence of how mouth-opening patterns and lateral motion vary across people. Our multi-sensor setup also allows us to identify which areas of the lower face offer the strongest biometric signals, opening the door to future optimizations.

Compared to previous approaches, our system supports a deeper and more targeted analysis of speech-related motion, grounded in the actual physical activity of speaking.

\paragraph{Resistance to Video-Based Attacks.} We evaluate a pure \emph{video-driven} impersonation threat in which the adversary collects public footage of the target, and, using face-tracking, converts landmark trajectories at our three sensor locations into synthetic 
\(X\), 
\(Y\), and 
\(Z\) accelerations.
Prior inertial-speech systems either omit this threat entirely~\cite{liu2021mandipass, liu2022secure, shi2021facemic, feng2017vauth} or address it only in passing.
The closest attempt is the work in~\cite{jawthenticate}, but that study blends its synthetic landmark motion with the attacker’s own inertial signals, leaving unclear how much of the final trace is truly derived from the video. It is also silent about key variables such as frame rate and resolution that heavily govern tracking fidelity in real-world videos. Moreover, their measurements come from sensors near the right ear, a modality distinct from our direct mouth-motion signals.

By focusing on pure video without any inertial input, explicitly studying video quality dynamics and doing these in the context of mouth movements, our study offers the first end-to-end assessment of how inertial mouth-movement biometrics withstand a video-only adversary.

\paragraph{Content-Independent Vibration Signatures.}
Our system supports \emph{free-speech} authentication, verifying a user during arbitrary conversation without requiring fixed prompts. By contrast, MandiPass~\cite{liu2021mandipass, liu2022secure} and Face-Mic~\cite{shi2021facemic} are built around short, prescribed utterances such as digits or common commands, and VAuth~\cite{feng2017vauth} limits users to a set of 30 predefined phrases. These schemes effectively learn a template for \emph{how the user pronounces a particular phrase} and then classify future attempts by matching that phrase-specific pattern. The approach cannot generalize if the speaker departs from the enrolled script. Jawthenticate~\cite{jawthenticate} does test free speech, but as noted earlier, their work studies a modality distinct from our direct mouth-motion patterns. 

\subsection{Other Approaches for Combating Fake Audio}

Since mitigating voice-based attacks is the primary motivation behind our approach, it is important to briefly situate our work within the broader landscape of techniques proposed to counter such threats.  

One of the most prominent approaches proposed to mitigate deepfake audio is watermarking. Watermarking works by embedding imperceptible signals into the audio waveform such that the presence of these signals can later be used to verify authenticity or origin. For example, in~\cite{roman2024proactive}, the authors present AudioSeal, which embeds localized, inaudible watermarks into synthetic speech and uses a fast detection model to highlight forged segments at a fine-grained level. WavMark~\cite{chen2023wavmark} is another technique which uses invertible neural networks to embed robust high-capacity watermarks that survive typical signal distortions and enable identification of AI-generated speech across long recordings.

Another family of techniques detects synthetic speech through learned heuristics. For instance, in~\cite{jung2022aasist}, the authors introduce AASIST, which uses a spectro-temporal graph attention network to capture subtle time-frequency artifacts produced by generative models. 

A third line of work explores digital signatures \cite{ssl}, a cryptographic technique that verifies the authenticity of the speaker or content origin. These systems tie audio content to a known identity using public-key infrastructure, such that any modification or impersonation can be detected. 

Our work can be viewed as a potential complementary technique to all of the above, particularly in scenarios where these methods face practical limitations. Watermarking may have limited impact given that open-source models released before watermarking standards emerged are already in circulation. Synthetic speech detectors must continually adapt to new forgery methods that remove the artifacts on which their models rely. Finally, the widespread application of digital signatures to all voice content published online would require major infrastructural changes, including scalable identity issuance and enforcement. Our scheme can operate in tandem with these approaches, reinforcing their strengths and offering an independent layer of defense. Moreover, it provides a usable fallback mechanism for securing voice-based interactions, even in the absence of any of the above protections.


\section{Data Collection and Classification System}
Following IRB approval, we conducted a series of experiments to evaluate the effectiveness of our system. This section describes our data collection process and the classification pipeline used to analyze the collected signals.
\begin{table}[ht]
\centering
\scriptsize
\caption{\textbf{User Demographic Distribution}}
\label{demographics}
\begin{tabular}{lrr}
\toprule
Language Background & Male & Female \\
\midrule
Native & 15 & 9 \\
Non-Native & 16 & 4 \\
\midrule
Total & 31 & 12 \\
\bottomrule
\end{tabular}
\end{table}


\begin{table*}[h]
\scriptsize
\centering

\caption{Top Performing Features Extracted from Various Sensor Locations Based on ReliefF Feature Importance Scores. Sensor Labels: C = Chin, LRC = Lower Right Cheek, ULC = Upper Left Cheek. X, Y and Z represent the axes from which the features are computed.}
\label{relieff-table}
\begin{tabular*}{\textwidth}{@{\extracolsep{\fill}} l r l r l r}
\toprule
\textbf{Feature (Sensor)} & \textbf{Score} & \textbf{Feature (Sensor)} & \textbf{Score} & \textbf{Feature (Sensor)} & \textbf{Score} \\
\midrule
Multiscale Entropy (LRC, Z) & 0.0358 & Centroid (LRC, X) & 0.0253 & Spectral Variation (ULC, Y) & 0.0216 \\
Detrended Fluctuation Analysis (LRC, Y) & 0.0354 & Average Power (LRC, Z) & 0.0252 & Mean Absolute Deviation (LRC, Y) & 0.0215 \\
Absolute Energy (ULC, Y) & 0.0346 & Median Absolute Deviation (C, Y) & 0.0250 & Histogram Mode (ULC, Z) & 0.0214 \\
Spectral Slope (C, Y) & 0.0345 & ECDF Slope (C, Z) & 0.0247 & Lempel-Ziv Complexity (ULC, X) & 0.0213 \\
Petrosian Fractal Dimension (ULC, Z) & 0.0342 & Signal Distance (ULC, X) & 0.0246 & Petrosian Fractal Dimension (LRC, Z) & 0.0213 \\
Histogram Mode (LRC, Z) & 0.0339 & Spectral Entropy (ULC, Z) & 0.0243 & Higuchi Fractal Dimension (C, Y) & 0.0211 \\
Spectral Entropy (C, X) & 0.0337 & Lempel-Ziv Complexity (ULC, Z) & 0.0241 & Human Range Energy (ULC, Y) & 0.0210 \\
Detrended Fluctuation Analysis (ULC, Y) & 0.0315 & Spectral Variation (ULC, Z) & 0.0240 & Median Absolute Deviation (LRC, Z) & 0.0210 \\
Spectral Skewness (ULC, Z) & 0.0289 & Spectral Kurtosis (LRC, Z) & 0.0240 & Spectral Positive Turning (ULC, Z) & 0.0210 \\
Sum of Absolute Differences (C, Y) & 0.0285 & Spectral Skewness (LRC, Z) & 0.0236 & Peak-to-Peak Distance (C, Y) & 0.0210 \\
Peak-to-Peak Distance (C, X) & 0.0282 & Sum of Absolute Differences (LRC, Z) & 0.0235 & Histogram Mode (LRC, Y) & 0.0203 \\
Spectral Roll-Off (ULC, Z) & 0.0281 & Wavelet Entropy (C, X) & 0.0233 & Wavelet Entropy (LRC, X) & 0.0203 \\
Lempel-Ziv Complexity (LRC, X) & 0.0271 & Higuchi Fractal Dimension (LRC, Y) & 0.0233 & Signal Distance (LRC, X) & 0.0199 \\
Histogram Mode (C, X) & 0.0271 & Absolute Energy (LRC, X) & 0.0232 & Signal Distance (C, Z) & 0.0197 \\
Spectral Skewness (C, Y) & 0.0269 & Minimum Value (ULC, Y) & 0.0230 & Centroid (ULC, X) & 0.0195 \\
Spectral Entropy (LRC, Y) & 0.0263 & Mean Absolute Difference (LRC, Z) & 0.0222 & Mean Absolute Deviation (ULC, X) & 0.0194 \\
Standard Deviation (LRC, X) & 0.0254 & & & & \\
\bottomrule
\end{tabular*}
\end{table*}


\subsection{Data Collection Process}
\label{datacoll}
We recruited 43 participants from our department, representing both undergraduate and graduate students with diverse linguistic backgrounds, including native and non-native English speakers. Of these participants, 31 were male and 12 were female, with ages ranging from 18 to 44 (see Table~\ref{demographics} for a detailed breakdown).

Each participant wore our prototype device and engaged in a free-form, interview-style conversation with a researcher. These sessions were designed to capture natural speech dynamics under different physical conditions.

Because the prototype made direct contact with the participant’s body, we followed hygiene protocols in line with IRB recommendations. Before each session, we sanitized all parts of the device, including the sensors, straps, and both the inner and outer surfaces of the tactical helmet, using disposable sanitary wipes~\cite{sanitaryWipes}. Participants also wore a disposable polypropylene cap~\cite{sanitaryCap} as a protective barrier beneath the helmet. After each session, the equipment was re-sanitized to maintain consistent hygiene across trials.

We conducted three types of experiments:

\textbf{\textit{Seated Conversation}}:
The participant was seated in a chair across from a member of our research team and engaged in a 15-minute open-ended conversation. This setup emulates typical speaking scenarios such as phone or video calls while seated. A total of 32 participants took part in this condition.

\textbf{\textit{Walking on a Flat Surface}}:
Participants wore the helmet while walking through a hallway in our department. A researcher walked alongside them, holding a continuous, casual conversation. This experiment tested the system’s robustness in motion, reflecting common situations like speaking on the phone while walking. Eleven participants completed this condition.

\textbf{\textit{Walking Upstairs}}:
Participants walked up a flight of stairs while conversing with a researcher. This condition was designed to evaluate how physical exertion and changes in elevation such as increased breathing or intermittent speech might affect speech dynamics. Speech during stair climbing is often more breathy or interspersed with pauses, making it a useful test for adaptability. Eleven participants completed this experiment.

Each participant completed two sessions per experiment type, spaced at least one day apart. One session was used for training and the other for testing. This temporal separation helped capture natural day-to-day variation in speech and behavior, providing a more realistic assessment of how our system would perform under real-world authentication scenarios.

\subsection{Classification Pipeline}

\subsubsection{Feature Extraction}
Using raw accelerometer data collected from the inertial sensor at each location, we extracted features using the TSFEL Python library \cite{tsfel}, a toolkit for time-series signal processing and feature engineering. A fixed window of 250 samples (approximately 2.5 seconds of data) was applied to segment the data before features were computed. Classification is performed at the window level. Longer utterances produce multiple windows, each classified independently, which naturally handles variability in utterance duration without padding or alignment.

From each window, a total of 54 features were computed from each of the X, Y and Z axes, creating a total of 162 features. The features spanned statistical, temporal, and spectral domains. Additional details on the implementation and descriptions of each feature can be found in the official TSFEL documentation \cite{tsfel}. We used the ReliefF feature selection algorithm \cite{kononenko1997overcoming} to select the 50 most powerful features out of this large feature-set. Table \ref{relieff-table} shows these top 50 feature along with their ReliefF scores.

\subsubsection{Classification Engines Used}
Throughout our experimentation phase, we explored various traditional machine learning classifiers and neural network models. Based on performance comparisons, we selected two standout models: one representing classical machine learning and the other representing deep learning. The classical model was a Support Vector Machine (SVM) utilizing a linear kernel with a C value of 1.0 and a fixed random state of 42 to ensure reproducibility across users. For the deep learning component, we employed Long Short-Term Memory (LSTM) networks to capture temporal dependencies within the accelerometer data and perform sequence-based classification. 

Our LSTM architecture is a two-layer design optimized for binary classification with dynamic learning rate adaptation and class imbalance mitigation. The initial LSTM layer uses 64 units, processes sequential input with return sequences to propagate temporal outputs to the subsequent layer, and applies dropout regularization of 0.3 to prevent overfitting. The second LSTM layer omits return sequences while maintaining identical unit counts and dropout configurations. For classification, we implemented a single-unit dense output layer with sigmoid activation. A learning rate of 5e-4 is used with Adam optimizer, binary cross-entropy loss function for the binary classification task and accuracy as our performance metric during training. Two callbacks are used to reinforce training robustness: Early stopping at 10 epochs and adaptive learning rate reduction of 0.2 with 5 epochs patience. To address class imbalance, instance weighting prioritizes minority-class samples, ensuring unbiased classification.

Note that because we are solving a biometric authentication problem, we build 2-class classifiers (SVM and LSTM) for each user to separate between the user's data and impostor data. The training datasets for both the SVM and LSTM models for each user were sourced solely from Session 1 of the experiment (recall experiment session in Section \ref{datacoll}). A 1.5 to 1 ratio of impostor to authentic user samples was used to introduce a mild class imbalance, which was found to enhance model performance. To preserve the independence between training and evaluation, the testing data was drawn exclusively from Session 2. The same 1.5 to 1 impostor to legitimate user ratio was applied to the testing set, with impostor data being sampled from all users except the user in question. Each model produced probabilistic outputs, which were used to calculate user-level Equal Error Rates (EER).  The final EER of the authentication system was computed as the median of the user-level EERs. This median statistic is stable and resists distortion from extreme EER values among subjects.

\subsection{Implementation of the Video-Driven Attack}

The impostor set described in the previous section represents a classic zero-effort attack. In this attack, samples from other users are simply matched against a target user's profile without any deliberate attempt to mimic the target. This type of attack is the standard baseline used to report the performance of biometric systems. However, a resourceful adversary is more likely to exploit publicly available video of the target.  Our video-driven attack experiments have three core operations: landmark tracking, quality-scale adjustment, and acceleration synthesis. 

\smallskip
\noindent\textbf{Landmark tracking.}
From a video clip that shows the target speaking, we track the three facial points corresponding to our sensor mounts (below the chin, left upper jaw, right lower jaw).  
Tracking uses Google MediaPipe \texttt{FaceMesh}~\cite{lugaresi2019mediapipe}, which returns 468 landmarks per RGB frame.  
\texttt{FaceMesh} provides pixel-normalised \(x,y\) directly and \emph{estimates} a relative \(z\) coordinate using a neural network trained on thousands of 3-D face scans.  
Because almost all Internet videos are monocular 2-D recordings, a real adversary would likewise lack true depth and be forced to rely on such an estimate.  Although the absolute camera-to-face distance is unknown, the inferred \(z\) is temporally consistent and thus sufficient for detecting frame-to-frame changes.

\smallskip
\noindent\textbf{Video-quality sweep.}
We capture the master video clip at \textbf{1080p 60 fps}, motivated by 
Full-HD (1080 p) being the most common resolution tier for user-generated video. Recording at the higher 60 fps, while still less prevalent than 30 fps, provides the best temporal detail from which we can cleanly derive lower frame rates.  
Each master clip is down-sampled with \texttt{FFmpeg} \cite{tomar2006converting} so that we ultimately have the following six quality levels:
\[
\{60,\,30,\,15\}\ \text{fps} \;\times\; \{1080p,\,720p\},
\]
These quality levels span everything from high-end smartphone footage to bandwidth-constrained webcam uploads, allowing our experiments to cover the full spectrum of video qualities an attacker is likely to encounter.

\smallskip
\noindent\textbf{Acceleration synthesis.}
For each quality level we obtain a time series \(\mathbf{p}_t=(x_t,y_t,z_t)^\top\).  
After anchoring all axes with a single pixel-to-metre factor, we compute linear acceleration via a second-order difference:
\[
\mathbf{a}_t
  =\frac{\mathbf{p}_{t+1}-2\mathbf{p}_t+\mathbf{p}_{t-1}}{\Delta t^{2}},\qquad
  \Delta t = 1/\text{fps}.
\]
where $t$ indexes video frames.
These synthetic \((a_{x,t},a_{y,t},a_{z,t})\) traces replace the inertial signals our hardware would have produced and are fed to the victim’s classifier to measure attack success as video quality degrades.

\begin{table*}[ht]
\centering
\scriptsize
\caption{User-level EER Distribution in Relation to Sensor Location. For each classifier and sensor placement, we report the percentage of speakers (N = 43) whose individual authentication model achieves an EER within the specified range. Lower ranges indicate better authentication performance.}
\label{EER-user-Distibution}
\begin{tabular}{@{}lcc|cc|cc|cc@{}}
\toprule
\diagbox[width=9em]{Sensor\\Location}{EER Range} 
& \multicolumn{2}{c|}{$<$ 0.05} 
& \multicolumn{2}{c|}{0.05–0.10} 
& \multicolumn{2}{c|}{0.10–0.30} 
& \multicolumn{2}{c}{$>$ 0.30} \\
\cmidrule(r){2-9}
& \textbf{LSTM} & \textbf{SVM} 
& \textbf{LSTM} & \textbf{SVM} 
& \textbf{LSTM} & \textbf{SVM} 
& \textbf{LSTM} & \textbf{SVM} \\
\midrule
Upper Left Cheek   & 71.9\% & 6.2\%  & 3.1\% & 12.5\% & 12.5\% & 56.2\% & 12.5\% & 25.0\% \\
Lower Right Cheek  & 81.2\% & 9.4\%  & 3.1\% & 3.1\%  & 6.2\%  & 62.5\% & 9.4\%  & 25.0\% \\
Below the Chin     & 78.1\% & 21.9\% & 3.1\% & 12.5\% & 9.4\%  & 34.4\% & 9.4\%  & 31.2\% \\
\bottomrule
\end{tabular}
\end{table*}


\section{Experimental Results and Evaluation}
\subsection{System Performance When Users Are Seated}

\subsubsection{Global System Performance}
Figure \ref{fused-eer-accents} shows the median EERs of our user population when users spoke while seated. The figure shows that the LSTM way outperforms the SVM when you consider the entire population and when you consider population segments of native speakers separately from non native speakers. The LSTM's EERs less than 0.01 for all scenarios are quite decent, especially  given the context of behavioral biometrics where error rates tend to be higher. 

As highlighted earlier, in practice we envision this kind of system to not necessarily interrupt the conversation when a classification errors happens. Rather, the errors can provide cues to asses the validity of a conversation after the fact if needed, so as to potentially pre-empt any fraudulent acts. So a single erroneous decision happening once in 100 window submissions should not be a problem for the user's experience. Also, note that the voice itself would be concurrently evaluated in real-time, providing complementary evidence to quickly ignore an erroneous inertial sensor classification if the voice in that particular segment is being verified with high confidence.  

Figure \ref{fused-eer-accents} shows that native speakers have lower median EERs than non-native speakers for both classifiers.
With the LSTM the difference is minimal because nearly all EER values are close to zero, whereas the SVM displays a clear and statistically significant gap.
A likely explanation is that non-native speakers introduce more articulatory variability, such as accent-related changes in mouth opening and timing. These factors might perhaps broaden the feature distribution and make it harder for certain classifiers to separate genuine and impostor samples. That said, larger and more diverse datasets will be necessary to definitely confirm whether language background systematically affects inertial mouth-movement biometrics.

\subsubsection{User-Level Performance}
The median EER used in the previous subsection has the advantage of being stable and resisting distortion from extreme values. This very strength of this statistic however also hides the users who fall in the long tail of the error distribution.  A small subset of speakers may still authenticate poorly even though the median suggests near-perfect performance. Examining user-level EERs therefore helps check if such users exist, allowing for possible discovery of any factors behind their performance. 

\begin{figure}
    \centering
    \includegraphics[width=.75\linewidth]{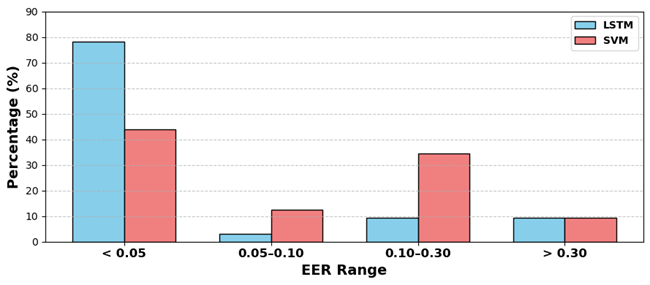}
    \caption{User-level EER Distribution in The Seated Condition. Each speaker is evaluated with an individual authentication model, and the bars show the percentage of speakers (N = 43) whose EER falls within each range. EER values are expressed as percentages.
}
    \label{FUSED-EER-user}
\end{figure}

Figure \ref{FUSED-EER-user} shows the complete distribution of user-level EERs. With both classifiers the vast majority of participants fall below the 0.05 threshold, which matches the strong system-wide results reported earlier. At the opposite extreme, about ten percent of the sample (three of the thirty two users) record EERs above 0.30 for both the LSTM and the SVM. These high-error cases raise the overall median. The same three individuals appear in the high-error tail for each model, and all three happen to be non-native English speakers.

\begin{figure}
    \centering
    \includegraphics[width=0.75\columnwidth]{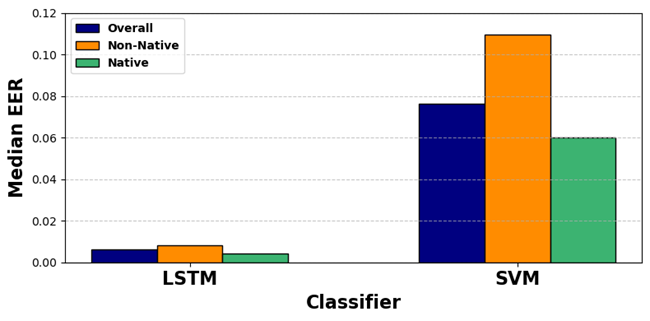}
    \caption{Median EER on Combined Locations for Users of Different Language Background in a Seated Conversation}
    \label{fused-eer-accents}
\end{figure}

It is noteworthy that the study includes seventeen other non-native speakers and that many of them appear in the lowest error bucket. This finding shows that nativeness alone cannot explain the high-error cases implied by Figure \ref{fused-eer-accents}. Other factors such as individual pronunciation habits, sensor fit, or speaking pace may be responsible for the elevated errors observed in those three users. Future work with a larger and more diverse cohort will be needed to pinpoint the variables that occasionally hinder authentication.

\subsubsection{Impact of Sensor Location}
\begin{figure}
    \centering
    \includegraphics[width=0.75\columnwidth]{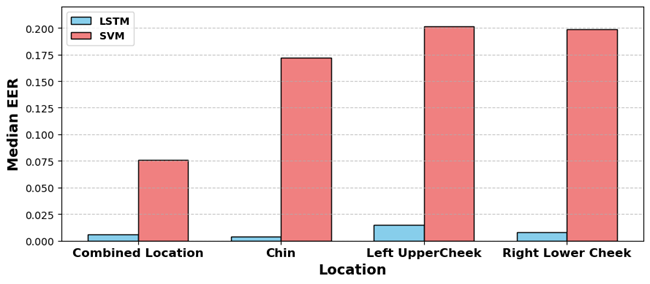}
    \caption{Median EER by Location for Seated Conversation Experiment}
    \label{locations}
\end{figure}

Table \ref{EER-user-Distibution} and Figure \ref{locations} compare each individual sensor with the fused three-sensor configuration. Figure \ref{locations} shows the global EERs, while Table \ref{EER-user-Distibution} breaks these down to a user-level. Across all experiments and for both classifiers, the placement directly beneath the chin consistently yields the lowest median EER, confirming our intuition that vertical jaw displacement is the single most informative cue for mouth kinematics. 

This chin placement is also unique to our design; none of the related works (see Section \ref{head-worn}) use a chin sensor. The two cheek sensors perform similarly to each other but contribute less discriminative power on their own. For the SVM, fusing all three sensors markedly lowers the median EER, indicating that the additional side information helps a margin-based model. For the LSTM, however, fusion provides little benefit beyond the chin sensor, suggesting that temporal modelling of the chin signal alone is sufficient.

From an engineering standpoint, a single sensor positioned beneath the chin can match the LSTM’s full three-sensor accuracy, lowering hardware cost, power draw, and wearer discomfort. For an SVM-based system the chin sensor still dominates, but fusing it with the two cheek sensors yields the lowest error, so a modest increase in complexity is justified if a margin-based classifier is preferred. On the scientific side these results underscore the chin’s importance in speech biomechanics: vertical jaw movement carries the most distinctive inertial signature for differentiating speakers, suggesting that future studies of mouth kinematics should treat this region as the primary source of biometric information.

\begin{table*}[h]
\scriptsize
\centering
\caption{Median EER Across User Activities}
\label{eer-table}
\begin{tabular*}{\textwidth}{@{\extracolsep{\fill}} l *{8}{c} }
\toprule
\diagbox[width=9em]{Experiment}{Location} 
& \multicolumn{2}{c}{\textbf{Fused}} 
& \multicolumn{2}{c}{\textbf{Below the Chin}} 
& \multicolumn{2}{c}{\textbf{Upper Left Cheek}} 
& \multicolumn{2}{c}{\textbf{Lower Right Cheek}} \\
\cmidrule(lr){2-3} \cmidrule(lr){4-5} \cmidrule(lr){6-7} \cmidrule(lr){8-9}
& \textbf{LSTM} & \textbf{SVM} 
& \textbf{LSTM} & \textbf{SVM} 
& \textbf{LSTM} & \textbf{SVM} 
& \textbf{LSTM} & \textbf{SVM} \\
\midrule
Seated Conversation       & 0.01 & 0.08 & 0.00 & 0.17 & 0.01 & 0.20 & 0.01 & 0.20 \\
Walking on a Flat Surface & 0.01 & 0.01 & 0.01 & 0.02 & 0.01 & 0.02 & 0.01 & 0.04 \\
Walking Upstairs           & 0.00 & 0.01 & 0.00 & 0.04 & 0.01 & 0.13 & 0.01 & 0.21 \\
\bottomrule
\end{tabular*}
\end{table*}

\subsection{System Performance When Users Are Walking}
Table \ref{eer-table} compares median EERs for three activities: seated conversation, walking on flat ground, and walking upstairs. The table also separates results by classifier and by sensor placement.

The LSTM remains highly accurate in every setting. Its median EER never rises above 0.01, and during stair climbing it falls to zero for both the fused configuration and the chin-only sensor, showing that the temporal model is almost unaffected by moderate body motion.

The SVM tells a more elaborate story. When users are seated the fused arrangement yields a respectable 0.08 EER, but individual cheek sensors reach 0.20. Once users begin to walk the SVM improves sharply: on flat ground the fused and chin configurations drop to 0.01–0.02. A likely reason is that head and jaw motion now incorporate the user's gait pattern, long known to be distinctive across individuals. This likely augments the speech-related features and gives the SVM clearer margins. The benefit persists for the chin sensor during stair ascent, yet the right-cheek sensor deteriorates to 0.21, suggesting that lateral strap movement introduces noise the margin-based model cannot reject.

Fusing all three sensors consistently helps the SVM, reducing its error by up to a factor of eight relative to the weakest location, whereas the LSTM gains little from fusion because the chin signal alone already captures the combined speech and gait dynamics.

In summary, vertical jaw motion recorded beneath the chin is the most stable and discriminative source across activities, and walking motion appears to enrich the signal for a classic SVM by layering user-specific gait information on top of speech kinematics.

\subsection{System Performance Under Video-Driven Attacks}
The attack is based on the scenario where users were seated. 
Each user’s \emph{EER threshold}, established from the zero-effort impostor evaluation that has been done for this scenario earlier in this section, serves as the decision boundary for this experiment.  
To test the video-driven attack, we feed the synthetic acceleration traces derived from public footage to the classifier trained for that user. We then determine whether an attack sample is successful or not by comparing its output with the user's threshold. 

We applied this procedure to five randomly selected users across the six video-quality settings (1080 p and 720 p at 60, 30, and 15 fps).  
Across all combinations—five users, two classifiers, six quality levels, and hundreds of attack windows, no attack sample successfully authenticated, giving a false-accept rate of 0 \%.

Several factors plausibly explain the attack’s failure.  
Because FaceMesh provides only a relative depth estimate inferred from a statistical shape model, small depth errors balloon into large acceleration errors after second-order differentiation.  
Most public clips are two-dimensional, so an attacker must contend with this limitation.  
Many articulatory motions, especially the subtle vertical movements of soft tissue beneath the chin, fall below pixel resolution even at 1080 p and are therefore lost or heavily quantised before differentiation.  
Finally, online frame rates of fifteen to sixty frames per second (far lower than the 100 Hz rate of our inertial sensors) miss high-frequency micro-vibrations, and differencing further amplifies this information gap.  Individually these issues are challenging; together they likely keep every synthetic trace far from the personalised threshold.  

These results indicate that inertial mouth-movement biometrics withstand a video-only adversary under realistic Internet video conditions.

\section{Discussion and Conclusions}
\label{sec:discussion}

This work set out to determine whether inertial mouth–motion patterns could serve as a practical second factor for voice authentication, whether the signal would remain robust in realistic usage, and whether publicly available video could be leveraged to defeat it.  
The evidence collected across 43 participants indicates that the answer to the first two questions is affirmative and to the last is, for now, negative.

\paragraph{What we learned.}  
A single IMU placed beneath the chin captures the essential dynamics of speech articulation.  
An LSTM fed only this signal achieves median EERs below $0.01$ while users are seated, walking, or climbing stairs.  
Cheek sensors add marginal benefit to a deep model but help a simpler SVM, confirming that lateral motion is informative even if it is secondary to vertical jaw displacement.  
Our video-driven attack, which combines state-of-the-art face tracking with aggressive frame-rate and resolution sweeps, never exceeded any user’s EER threshold.  
Depth ambiguity, sub-pixel motion, and limited temporal resolution appear to combine in a way that thwarts a video-only adversary.

\paragraph{Engineering implications.}  
These results suggest a streamlined hardware path: a lightweight chin-mounted module, powered by a small battery and sampling at 100 Hz, is enough for deep-learning deployments.  
In contexts where on-board processing power is scarce, two cheek sensors can be retained and fused with a margin-based classifier to keep error rates low.  
Either configuration demands far less equipment than headset-based systems explored in earlier work.

\paragraph{Scientific insight.}  
The study confirms what speech scientists have known for years: the up-and-down movement of the lower jaw is the main physical motion we make when we talk.
It also hints that whole-body rhythm, visible in the SVM’s improved accuracy while users walk, reinforces lower-face kinematics in biometric systems such as ours, an intersection worth probing with larger cohorts.

\paragraph{Applications and adoption.}  
Several domains already require head-mounted gear, making integration straightforward.  
Military helmets, firefighter masks, industrial hard hats, and aviation headsets could host a thin chin sensor with negligible extra burden, adding a real-time check against voice spoofing that could prevent catastrophic orders or miscommunication.  
In high-value financial, legal, or diplomatic calls, executives might accept a discreet chin strap or adhesive patch when authorising large transactions.  
Telemedicine platforms and emergency-response centres could deploy the sensor on top of existing protective equipment, gaining continuous speaker assurance without disrupting workflow.  
Where headgear is not yet routine, the security gains demonstrated here offer a concrete incentive to adopt minimal, user-friendly designs such as transparent chin patches.

\paragraph{Limitations and future work.}

Despite encouraging accuracy and observed robustness to video driven forgery under realistic Internet video conditions, several limitations remain.
First, the study cohort (43 participants) is modest and demographically unbalanced by gender, native versus non native status, and age. Consequently, subgroup trends, such as native versus non native comparisons, and sensor ablation conclusions, such as the observation that the chin signal can match fused performance, should be interpreted as preliminary rather than universal. A larger and more balanced dataset spanning broader age ranges, accents and language backgrounds, and additional populations will be necessary to validate these effects and to reassess the relative contribution of the cheek sensors with stronger statistical support.

Second, our current evaluation trains a participant specific model and evaluates it on a separate session from the same participant. This reflects a plausible personal device setting, but it does not capture deployment paradigms in which users are unseen during model development and are enrolled with one or more templates. Future work will therefore examine enrollment based protocols with unseen users, including template based enrollment, and longitudinal template updates.

Third, while our video only adversary fails, our attack model reflects what is feasible under current data availability, where paired inertial signals and corresponding audio or video are not publicly available at scale. In future, when such datasets become available at scale, it is arguable that a strong adversary could train a generative model to synthesize inertial mouth motion traces conditioned on speech and/or video, potentially supporting zero shot or few shot speaker adaptation. Quantifying this risk calls for new multimodal corpora and systematic evaluation of learned forgeries, including studies using higher frame rate and higher resolution capture when available. On the defense side, we plan to explore challenge response prompts and cross modality consistency checks that bind inertial dynamics to the spoken content. Finally, comfort, aesthetics, and long term wearability remain important open questions, and will be examined alongside accuracy and security in future work.

\paragraph{Conclusion.}  
Voice deepfakes are becoming a driver of social-engineering attacks, yet voice alone offers no easy second factor. 

Inertial measurements of mouth motion might provide that missing layer.  
They are accurate, resilient to everyday movement, and, at current video quality levels, resistant to optical imitation.  
With modest hardware and well-understood machine-learning models, the technique can harden many critical voice channels, giving defenders time to stay ahead of the next wave of synthetic speech.

\bibliographystyle{IEEEtran}
\bibliography{References}

\section*{Acknowledgment}
This work was supported by the U.S. Department of Defense Army Research Lab under Contract No. CA W911NF-24-2-0180. The opinions, findings, and conclusions or recommendations expressed herein are those of the authors and do not necessarily reflect the views of the U.S. Department of Defense.
\vspace{-2em}

\begin{IEEEbiography}[{\includegraphics[width=1in,height=1.25in,clip,keepaspectratio]{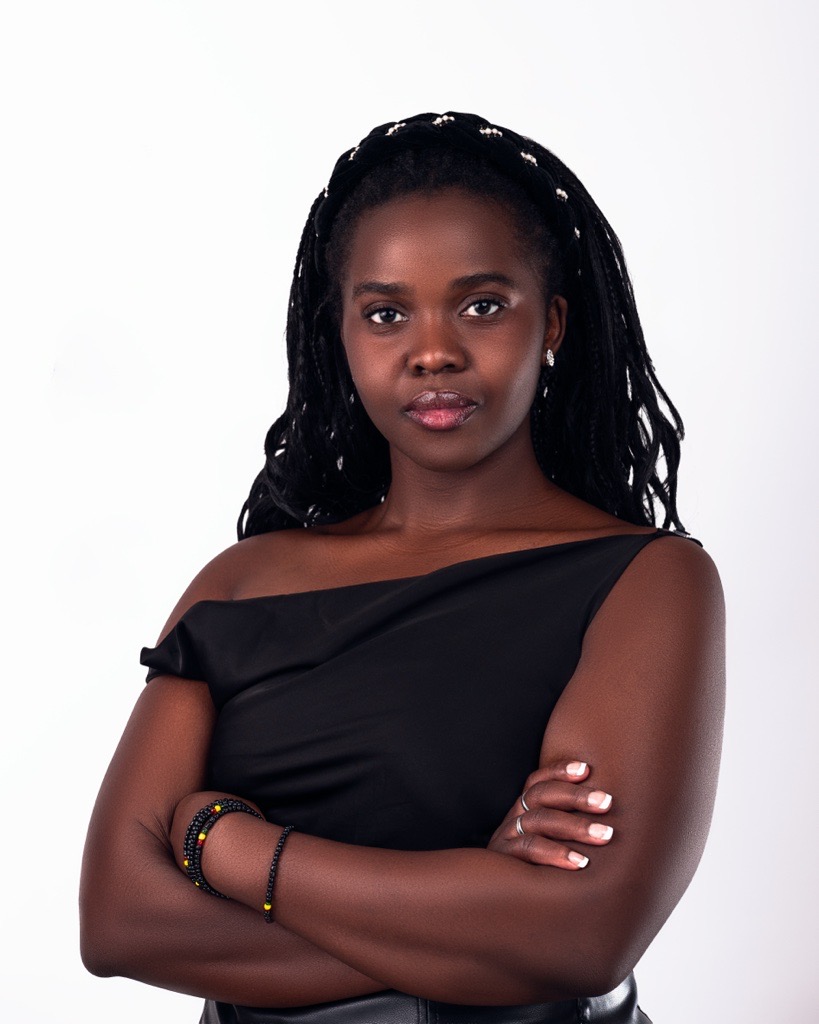}}]{Ynes Ineza} received her B.S. degrees in Mathematics and Computer Science from Texas Tech University in 2024. She is currently pursuing the Ph.D. in in Computer Science at Texas Tech University. Her research interests include file-encrypting ransomware detection and behavioral biometrics.
\end{IEEEbiography}

\begin{IEEEbiography}[{\includegraphics[width=1in,height=1.25in,clip,keepaspectratio]{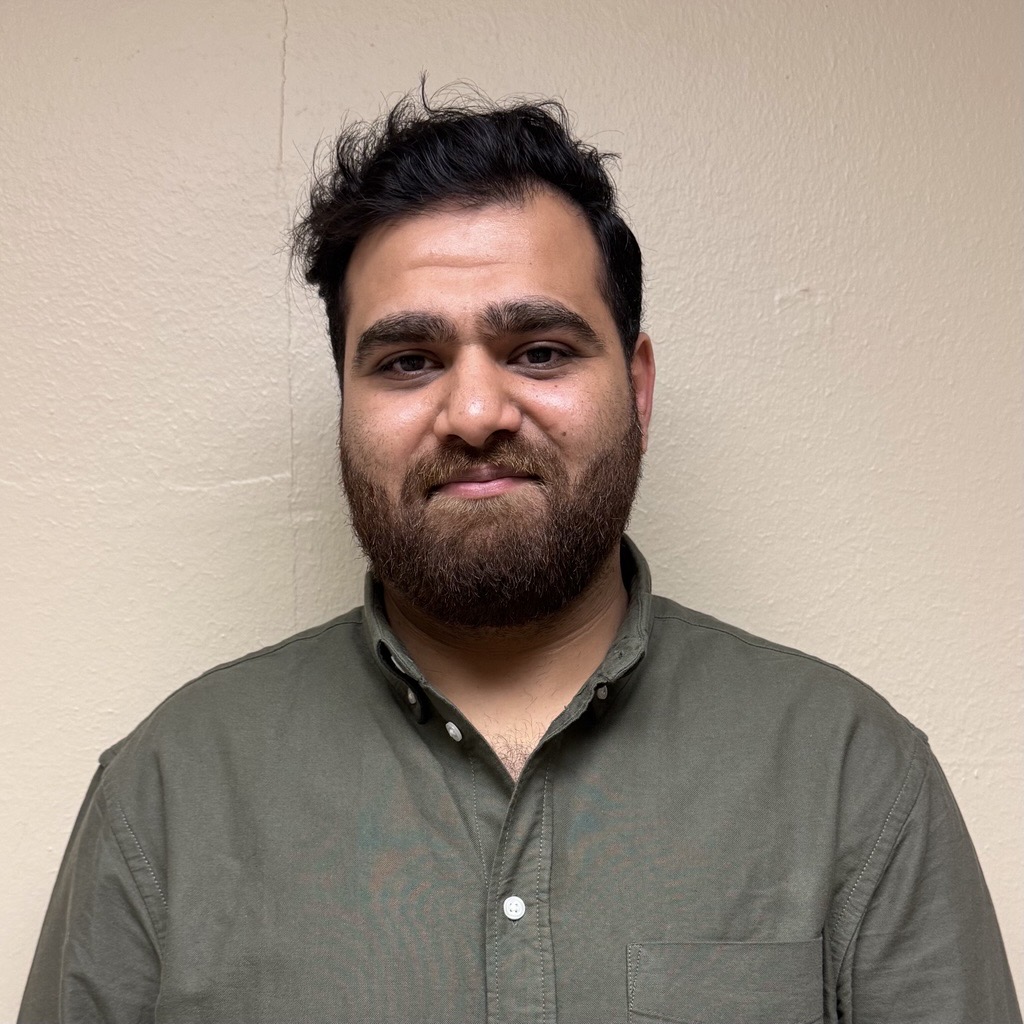}}]{Muhammad Aziz Ullah}
 received the B.S. degree in computer science from the National University of Computer and Emerging Sciences (FAST NUCES), Lahore, Pakistan, in 2020. He is currently working toward the Ph.D. degree in computer science with a concentration in cybersecurity at Texas Tech University, Lubbock, TX, USA.
He is currently a Research Assistant with Texas Tech University, Lubbock, TX, USA. His research interests include cybersecurity, artificial intelligence, large language models (LLMs), AI agents, and LLM security.
\end{IEEEbiography}
\begin{IEEEbiography}[{\includegraphics[width=1in,height=1.25in,clip,keepaspectratio]{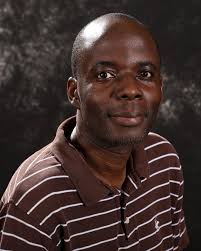}}]{Abdul Serwadda}
is an Associate Professor of Computer Science at Texas Tech University.  He received M.S. degrees in Computer Science and Mathematics and the Ph.D. degree from Louisiana Tech University. He also worked as a Post-Doctoral Fellow with Louisiana Tech and Syracuse Universities. His research is at the intersection of Cybersecurity and Artificial Intelligence. 

\end{IEEEbiography}

\begin{IEEEbiography}[{\includegraphics[width=1in,height=1.25in,clip,keepaspectratio]{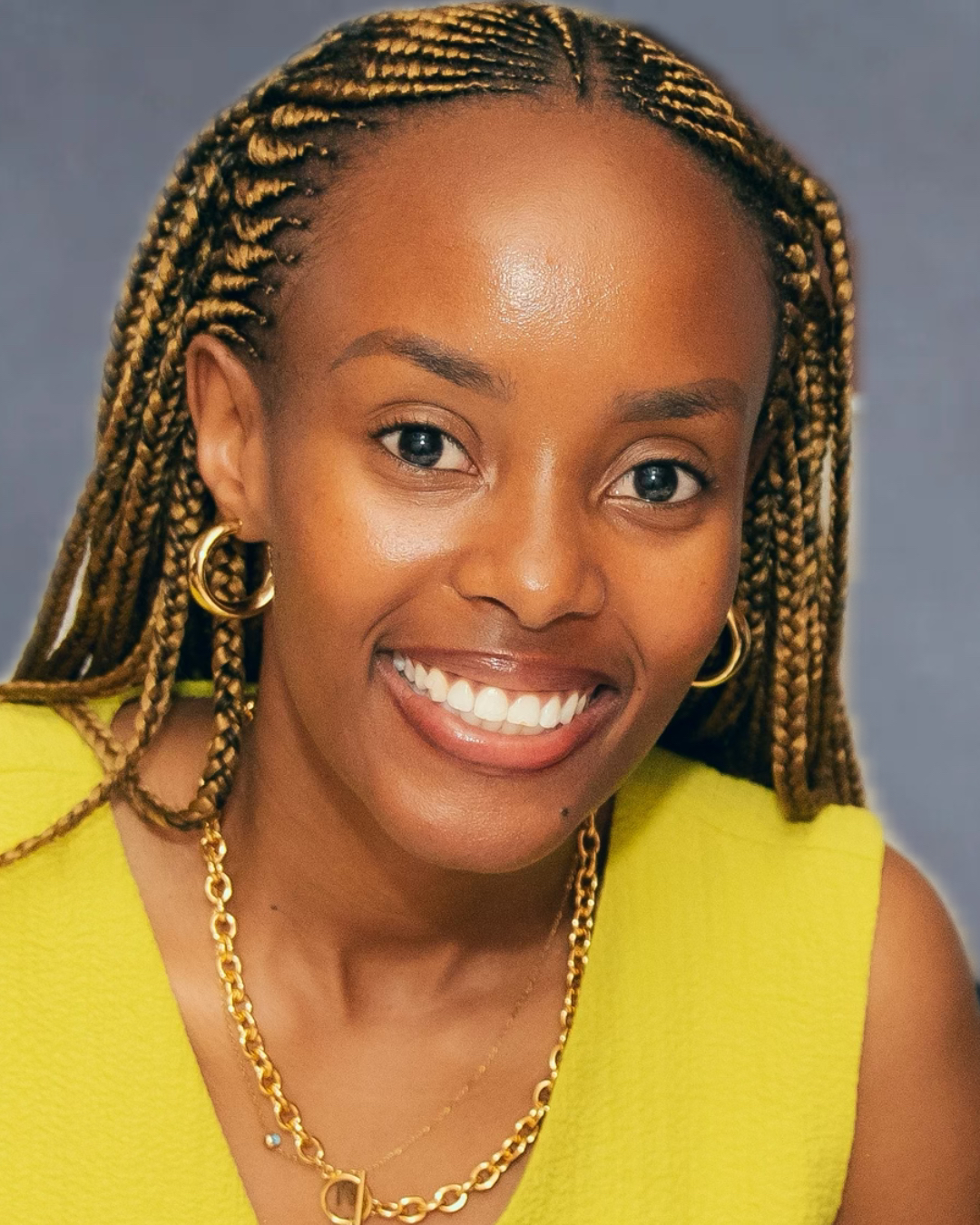}}]{Aurore Nicole Munyaneza}
received the B.Sc. degree in Mathematics with a minor in Computer 
Science and the M.Sc. degree in Computer Science from Texas Tech University, Lubbock, TX, 
USA. She has made research contributions in biometrics, including work on high-security speech 
verification using inertial sensing of mouth motion, and in user trust and privacy in mobile 
applications, focusing on the effectiveness of privacy labels. Her research interests include 
artificial intelligence and machine learning, biometric systems, and biomedical science.
\end{IEEEbiography}

\end{document}